\documentclass[%
reprint,
superscriptaddress,
 amsmath,amssymb,
 aps, physrev,
]{revtex4-2}

\usepackage{graphicx}
\usepackage{dcolumn}
\usepackage{bm}

\begin{document}

\preprint{APS/123-QED}

\title{Simple mathematical model for a pairing-induced motion of active and passive particles}%

\author{Hiroaki Ishikawa}
\affiliation{
 Department of Physics, Chiba University, 1-33 Yayoi-cho, Chiba 263-8522, Japan
}
\author{Yuki Koyano}
\affiliation{Department of Human Environmental Science, Graduate School of Human Development and Environment, Kobe University, Kobe 657-0011, Japan}
\author{Hiroaki Ito}
\affiliation{
 Department of Physics, Chiba University, 1-33 Yayoi-cho, Chiba 263-8522, Japan
}
\author{Yutaka Sumino}
\email{ysumino@rs.tus.ac.jp}
\affiliation{Department of Applied Physics, Tokyo University of Science, 6-3-1 Niijuku, Katsushika, Tokyo, 125-8585, Japan}
\affiliation{Water Frontier Science \& Technology Research Center, and Division of Colloid and Interface Science, Research Institute for Science \& Technology, Tokyo University of Science, 6-3-1 Nijuku, Katsushika-ku, Tokyo, 125-8585, Japan
}%
\author{Hiroyuki Kitahata}
\email{kitahata@chiba-u.jp}
\affiliation{
 Department of Physics, Chiba University, 1-33 Yayoi-cho, Chiba 263-8522, Japan
}
\date{\today}

\begin{abstract}
We propose a simple mathematical model that describes a pairing-induced motion of active and passive particles in a two-dimensional system, which is motivated by our previous paper [Ishikawa et al., Phys. Rev. E \textbf{106} (2022) 024604]. We assume the following features; the active and passive
particles are connected with a linear spring, the active
particle is driven in the direction of the current velocity, and the passive particle is repelled from the active particle. A straight motion, a circular motion, and a slalom motion were observed by numerical simulation. Theoretical analysis reproduces the bifurcation between the straight and circular motions depending on the magnitude of self-propulsion.
\end{abstract}

\maketitle

\section{\label{sec:introduction}Introduction}

Particles can exhibit self-propulsion by consuming free energy in nonequilibrium systems~\cite{Bechinger_RMP,Michelin_AnnuRev}. Some self-propelled particles can move through momentum exchange with surrounding fluid and they are often called as swimmers~\cite{toyota2009JACS,Thutupalli_2011,Yabunaka_Ohta_Yoshinaga_2012,Yoshinaga_PRE_2012,Yoshinaga_PRE2014,PhysRevLett_Izri,Herminghaus_SoftMatter2014,Maass,yamamoto_SoftMatter,Suga_Kimura_2018}. Other particles move due to the surface tension gradient at the liquid surface and they are called as Marangoni surfers~\cite{Nakata_camphor,Nagayama_PhysicaD,Kitahata_PhysicaD2005,Nagai_PRE2005,Lauga_Davis_2012,Boniface_PRE,PCCP_camphor_review2015,Gouiller_PRE2021}. Cells on the substrate also exhibit self-propulsion by the momentum exchange with the substrate~\cite{Bosgraaf_PlosONE2009,Levine_PRL2010,Taniguchi_Sawai_PNAS,Sens_PNAS,Tarama_JPSJ}. 
The collective behavior of these self-propelled particles has also attracted intensive interest for their rich variety of macroscopic dynamics.~\cite{SK005,SK006,Viscek_2012,SK049,annurev_Chate,Suzaka_PRE}. In many cases, the homogeneous systems comprising equivalent self-propelled particles have been studied. However, in actual systems, the heterogeneity of the comprising self-propelled particles often plays an essential role~\cite{Cira2015,Zarzar2020,KOJIMA2023,Ishikawa_PRE}. 

In the last decade, multiple-particle systems comprising heterogeneous self-propelled particles have been studied, in which the emergence of characteristic spatio-temporal patterns was reported~\cite{Soto_Golestanian_PRL2014,Soto_Golestanian_PRE_2015,Menzel_PhysRevE.93.022610,Golestanian2019,Gupta_PRE2022,Kreienkamp_2022,Dinelli2023,Zhang_PRR2023}. 
For example, Agudo-Canalejo and Golestanian reported the spatio-temporal pattern formation with phase separation of the two species interacting through nonreciprocal interaction~\cite{Golestanian2019}. Even a system comprising a single pair of two species is known to exhibit novel spatio-temporal behavior, pairing-induced motion. For example, Cira et al.~reported a pair of propylene glycol aqueous droplets with different concentrations exhibits a pairing-induced motion on a solid substrate~\cite{Cira2015}, and Meredit et al.~reported a pair of a fluorine oil and a carbon oil shows a pairing-induced motion in an aqueous solution through nonreciprocal interaction~\cite{Zarzar2020}. Kojima et al.~reported that oil droplets with different sizes form a pair and moves in an aqueous solution of photosensitive surfactants~\cite{KOJIMA2023}. K\"{u}chler et al. reported that active and passive beads connected by a linear spring under the external flow exhibit a non-trivial motion~\cite{Menzel_PhysRevE.93.022610}.

Recently, we reported that a pair of the source and inert particles show straight or rotational motions depending on the parameters by experiments and numerical simulations~\cite{Ishikawa_PRE}. In this system, we considered a concentration field of a chemical compound that is released from the source particle and the chemical concentration gradient drives both the particles. This system can be realized as an actual experimental system by using a camphor disk and a metal washer floating at a water surface. The pair exhibited a straight motion when the resistance was greater while it exhibited a circular motion when the resistance was smaller. The transition between these motions is understood through the linear stability analysis of our mathematical model. However, the physical mechanism of the pairing-induced motion and the transition between the modes of motion is still unclear. Motivated by our previous work~\cite{Ishikawa_PRE}, we construct a simple mathematical model that can extract the essence of pairing-induced motions.

In the present study, we propose a simple mathematical model for a pair of active (source) and passive (inert) particles in a two-dimensional system. In our model, we assume the following features; the active and passive particles are connected with a linear spring, the active particle is driven in the direction of its current velocity, and the passive particle is repelled from the active particle. By numerical simulations, we show the three characteristic motions: a straight motion, a circular motion, and a slalom motion. Then we perform the linear stability analysis and discuss the bifurcation between these motions. In Sect.~\ref{sec:model}, we introduce the model for the active and passive particles. The results of numerical simulations and linear stability analyses are shown in Sect.~\ref{sec:simulation} and Sect.~\ref{sec:analysis}. Finally, we discuss the results in Sect.~\ref{sec:discussion} and summarize our work in Sect.~\ref{sec:summary}.

\section{\label{sec:model}Model}

\begin{figure}[t]
\centering
\includegraphics{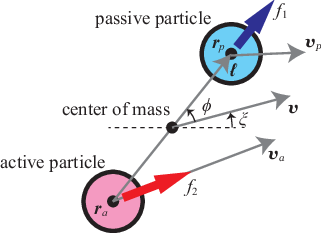}
\caption{(Color online) Schematic illustration of our model. An active particle and a passive particle are connected with a linear spring. Additionally, the passive particle receives the force $f_1$ in the direction of the vector $\bm{e}_{ap}$ directing from the active particle to the passive particle. The active particle receives the force $f_2$ in its moving direction. The angles $\xi$ and $\phi$ show the direction of the center-of-mass (COM) velocity $\bm{v}$ and the angle between the two vectors $\bm{\ell}$ and $\bm{v}$.
\label{fig1}}
\end{figure}

We consider the active and passive particles constrained at a two-dimensional surface. Their center positions are defined as $\bm{r}_a$ and $\bm{r}_p$, respectively. The velocities of them are also set to be $\bm{v}_a$ and $\bm{v}_p$.
We consider the linear spring that connects the active and passive particles, whose spring constant and natural length are $k$ and $R$, respectively. We assume that a constant driving force is applied on the active particle in the direction of its own velocity, and a constant non-reciprocal repulsive force is applied on the passive particle in the direction from the active particle to the passive one. The masses and the resistance coefficients of the particles are commonly set to $m$ and $\eta$ for both. The equations of motion can be explicitly described as
\begin{align}
\frac{d \bm{r}_p}{dt} = \bm{v}_p,
\end{align}
\begin{align}
\frac{d \bm{r}_a}{dt} = \bm{v}_a,
\end{align}
\begin{align}
    m \frac{d \bm{v}_p}{dt} = - \eta \bm{v}_p - \left [ k \left( \left| \bm{r}_p - \bm{r}_a \right| - R \right) - f_1 \right ] \bm{e}_{ap}, \label{eq_vp}
\end{align}
\begin{align}
    m \frac{d \bm{v}_a}{dt} = - \eta \bm{v}_a + f_2 \frac{\bm{v}_a}{|\bm{v}_a|} + k \left( \left| \bm{r}_p - \bm{r}_a \right| - R \right) \bm{e}_{ap},
\label{eq_va}
\end{align}
where $f_1$ and $f_2$ are the magnitude of the non-reciprocal repulsive force working on the passive particle and that of the self-propulsion force working on the active particle, respectively. $\bm{e}_{ap}$ is the unit vector in the direction from the active particle to passive one, i.e., $\bm{e}_{ap} = \left( \bm{r}_p - \bm{r}_a \right) / \left| \bm{r}_p - \bm{r}_a \right|$. The schematic illustration for our model is shown in Fig.~\ref{fig1}. 

This situation captures the essential behavior of the mathematical model for particles interacting through chemo-repellant concentration fields proposed in our previous paper~\cite{Ishikawa_PRE}, where the source (active) and inert (passive) particles for the concentration field is included. The source particle emits the chemo-repellant, which leads the self-propulsion force denoted by $f_2$ and the nonreciprocal interaction to the inert particle denoted by $f_1$. In the proposed model, an additional attractive interaction is also included in order to form a particle pair, which is simplified as a linear spring term in Eqs.~\eqref{eq_vp} and \eqref{eq_va}. The model in our previous paper~\cite{Ishikawa_PRE} was originally intended to reproduce the motion of a pair of a camphor disk and a metal washer floating on water, in which camphor molecules work as a chemo-repellant and the lateral capillary interaction works as an attractive interaction. However, the model can apply to general pairing-induced motions with an attractive potential force under a chemo-repellant concentration field.

Hereafter, we consider the dimensionless version of our model, in which the units of mass, length, and time are set to be $m$, $R$, and $\sqrt{m/k}$, respectively. Then, the time evolution equations are rewritten as
\begin{align}
\frac{d \bm{r}_p}{dt} = \bm{v}_p,
\end{align}
\begin{align}
\frac{d \bm{r}_a}{dt} = \bm{v}_a,
\end{align}
\begin{align}
    \frac{d \bm{v}_p}{dt} = - \eta \bm{v}_p - \left [ \left( \left| \bm{r}_p - \bm{r}_a \right| - 1 \right) - f_1 \right ] \bm{e}_{ap},
\end{align}
\begin{align}
    \frac{d \bm{v}_a}{dt} = - \eta \bm{v}_a + f_2 \frac{\bm{v}_a}{|\bm{v}_a|} + \left( \left| \bm{r}_p - \bm{r}_a \right| - 1 \right) \bm{e}_{ap}.
\end{align}
Here, $\eta/ \sqrt{mk}$, $f_1/ k$, and $f_2/k$ in the dimensional model are rewritten as $\eta$, $f_1$, and $f_2$, respectively. We consider them to be the essential parameters in the dimensionless model. From the correspondence to the considered situation, these three parameters should have positive values.

The evolution equations become easier to understand if we introduce the center-of-mass (COM) position $\bm{r} = \left(\bm{r}_a + \bm{r}_p\right)/2$ and the relative position $\bm{\ell} = \bm{r}_p - \bm{r}_a$. Accordingly, we introduce the COM velocity $\bm{v} = \left(\bm{v}_a + \bm{v}_p\right)/2$ and the relative velocity $\bm{w} = \bm{v}_p - \bm{v}_a$.
Then, our model is described as
\begin{align}
\frac{d \bm{r}}{dt} =& \bm{v}, \label{eq_r}
\end{align}
\begin{align}
\frac{d \bm{\ell}}{dt} =& \bm{w}, \label{eq_ell}
\end{align}
\begin{align}
\frac{d \bm{v}}{dt} =
 - \eta \bm{v} + \frac{f_1}{2} \frac{\bm{\ell}}{|\bm{\ell}|} + \frac{f_2}{2} \frac{\bm{v} - \bm{w}/2}{|\bm{v} - \bm{w}/2|}, \label{eq_v}
\end{align}
\begin{align}
    \frac{d \bm{w}}{dt} = - \eta \bm{w} + f_1 \frac{\bm{\ell}}{|\bm{\ell}|} - 2 ( |\bm{\ell}| - 1 ) \frac{\bm{\ell}}{|\bm{\ell}|} - f_2 \frac{  \bm{v} - \bm{w}/2 }{|\bm{v} - \bm{w}/2|}.  \label{eq_w}
\end{align}
It should be noted that our system is essentially a six-dimensional dynamical system since $\bm{r}$ is not included in the righthand sides in Eqs.~\eqref{eq_ell}, \eqref{eq_v}, and \eqref{eq_w}. Equations~\eqref{eq_v} and \eqref{eq_w} have singularities at $\left| \bm{\ell}\right| = 0$ and $\bm{v} = \bm{w}/2$. $\left| \bm{\ell} \right| = 0$ corresponds to the case that the active and passive particles are located at the same position, and thus we exclude the case with $\left| \bm{\ell} \right| = 0$. $\bm{v} = \bm{w}/2$ corresponds to $\bm{v}_a = \bm{0}$, which can be excluded since the active particle is always driven by the self-propulsive force.

\section{\label{sec:simulation}Numerical simulation}

In order to investigate the behavior of the system comprehensively, we performed numerical simulation based on the model described in the previous section. The numerical simulation was performed by the 4th-order Runge-Kutta method with a time step $\Delta t = 0.0001$ until $t = 10000$. The initial condition was set as $\bm{r}_p = 0.5 \bm{e}_x$, $\bm{r}_a = -0.5\bm{e}_x$, $\bm{v}_p = \bm{e}_x + 0.1 \bm{e}_y$, and $\bm{v}_a = 1.1\bm{e}_x + 0.05\bm{e}_y$ unless otherwise commented. The parameter $\eta$ was fixed at $\eta = 0.5$ since we confirmed that similar tendencies were observed for the other values for $\eta$.

\begin{figure}
\centering
\includegraphics{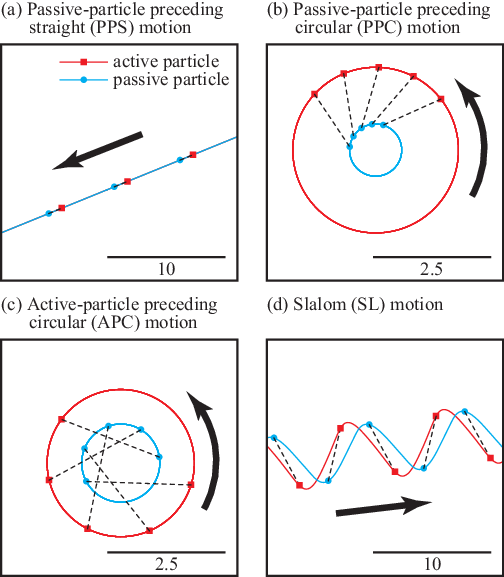}
\caption{(Color online) Trajectories and the instant locations of active and passive particles after sufficient time for four characteristic modes of motion. (a) Passive-particle preceding straight (PPS) motion. (b) Passive-particle preceding circular (PPC) motion. (c) Active-particle preceding circular (APC) motion. (d) Slalom (SL) motion. Red and cyan curves show the trajectories of active and passive particles, respectively. The parameters were set to be (a) $f_2 = 0.1$, (b) $f_2 = 0.3$, (c) $f_2 = 1.0$, and (d) $f_2 = 2.5$. The parameters $f_1$ and $\eta$ were fixed at $f_1 = 0.5$ and $\eta = 0.5$, respectively. The dashed lines show the correspondence of active and passive particle positions at each instance taken every (a) 10, (b) 1, (c) 1, and (d) 3 time unit. The bar in each panel shows the spatial unit. \label{fig2}}
\end{figure}

In Fig.~\ref{fig2}, we exhibit the trajectories and instant locations of active and passive particles for four characteristic types of motion. The pair of active and passive particles exhibited a straight motion with a passive particle in the front: passive-particle preceding straight (PPS) motion as shown in Fig.~\ref{fig2}(a). The trajectories of both particles coincided. As for the circular motion, there were two types: the passive-particle preceding circular (PPC) motion as shown in Fig.~\ref{fig2}(b) and the active-particle preceding circular (APC) motion as shown in Fig.~\ref{fig2}(c). For the circular motion, the radius of the trajectory of the active particle was always greater in both cases. The slalom (SL) motion, in which the active particle preceded, was also observed as shown in Fig.~\ref{fig2}(d). The trajectory of the passive particle was slightly shifted in the traveling direction, and the amplitude of the waving trajectory was greater for the active particle.

\begin{figure}
\centering
\includegraphics{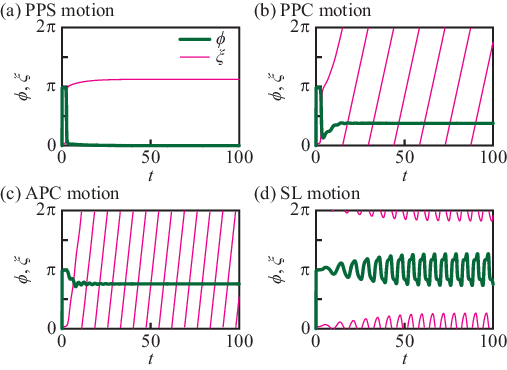}
\caption{(Color online) Time series of $\phi$ and $\xi$ at (a) $f_2 = 0.1$ for the PPS motion, (b) $f_2 = 0.3$ for the PPC motion, (c) $f_2 = 1.0$ for the APC motion, and (d) $f_2 = 2.5$ for the SL motion. Each panel corresponds to Fig.~\ref{fig2}. The parameters $f_1$ and $\eta$ were fixed at $f_1 = 0.5$ and $\eta = 0.5$, respectively. The dark green thick curve shows $\phi$, which is the angle between the vector directing from the active particle to the passive particle $\bm{\ell} = \bm{r}_p - \bm{r}_a$ and the COM velocity vector $\bm{v}$. The magenta thin curve shows $\xi$, which is the direction of COM velocity $\bm{v}$. \label{fig3}}
\end{figure}

In Fig.~\ref{fig3}, we show the time series of the angle difference $\phi \in [0, 2\pi)$ between the COM velocity $\bm{v}$ and the vector directing from the active particle to the passive particle $\bm{\ell}$. We also show the angle $\xi \in [0, 2\pi)$, which represents the direction of the COM velocity $\bm{v}$. In the case of the PPS, $\phi$ converged to 0 as shown in Fig.~\ref{fig3}(a). When the pair exhibited the circular motion, $\phi$ converged to a finite value between $0$ and $\pi$. If the converged value was smaller than $\pi/2$ as shown in Fig.~\ref{fig3}(b), then the passive particle preceded, i.e., the PPC motion was exhibited. In contrast, if the converged angle of $\phi$ was greater than $\pi/2$ as shown in Fig.~\ref{fig3}(c), the active particle preceded, i.e., the APC motion was exhibited. The value of $\phi$ oscillated around $\pi$ when the pair showed SL motion, as shown in Fig.~\ref{fig3}(d).

\begin{figure}[t]
\centering
\includegraphics{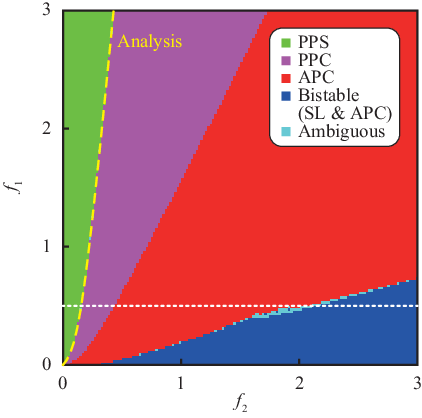}
\caption{(Color online) Phase diagrams in the $f_1$-$f_2$ plane when $\eta$ was fixed at $\eta = 0.5$. Green, purple, red, blue, and cyan regions show the PPS motion, the PPC motion, the APC motion, bistability between the SL and the APC motions, and the ambiguous motion, respectively. The white dotted line corresponds to the situation shown in Fig.~\ref{fig5}. The yellow dashed line shows the bifurcation line obtained by the theoretical analysis in Eq.~\eqref{line_th}.
\label{fig4}}
\end{figure}

The phase diagrams were obtained in the $f_1$-$f_2$ plane as shown in Fig.~\ref{fig4}. Based on the time series shown in Fig.~\ref{fig3}, we categorized the pair-induced motion as follows: First, we obtained the maximum and minimum values, $\phi_\mathrm{max}$ and $\phi_\mathrm{min}$, of $\phi$ and the maximum and minimum values, $C_\mathrm{max}$ and $C_\mathrm{min}$, of $\cos \phi$ during the time range $9500 \leq t \leq 10000$. If $C_\mathrm{min} > 1 - \epsilon$, then the motion was categorized into the PPS motion. If $C_\mathrm{max} - C_\mathrm{min} < \epsilon$ and $\xi$ sweeps $0$ to $2\pi$, then the motion is categorized into the circular motion. In this case, the motion is further categorized into two cases; the PPC motion if $\left(\phi_\mathrm{max} + \phi_\mathrm{min} \right)/2 < \pi/2$ and the APC motion if $\left(\phi_\mathrm{max} + \phi_\mathrm{min} \right)/2 \geq \pi/2$. The SL motion is characterized by the oscillation of $\phi$ around $\pi$. Therefore, the motion is categorized into the slalom motion if $\pi - \epsilon < (\phi_\mathrm{max} + \phi_\mathrm{min})/2 < \pi + \epsilon$. Here, we set $\epsilon = 0.0001$. If the motion was categorized into none of the above-mentioned motions, then the motion was defined as the ambiguous motion.

In order to check the bistability, we scanned the parameters $f_1$ and $f_2$ with the interval of 0.02 both in the upward and downward directions. The final values of $\bm{r}_a$, $\bm{r}_p$, $\bm{v}_a$, and $\bm{v}_p$ in the previous calculation were adopted as the initial condition of the next calculation after adding small random values, which obey the uniform distribution ranging $[-10^{-4}, 10^{-4}]$. The initial conditions for the first calculation were set as mentioned above.

As seen in Fig.~\ref{fig4}, the PPS motion as in Fig.~\ref{fig2}(a) was realized for the small ratio of $f_2/f_1$. As $f_2/f_1$ increased, the PPC motion as in Fig.~\ref{fig2}(b) was observed, and then the APC motion as in Fig.~\ref{fig2}(c) was observed. With the further increase in $f_2/f_1$, the SL motion, in which the active particle preceded and the passive particle followed, as in Fig.~\ref{fig2}(d), was observed. In this region, the APC motion was also observed depending on the initial condition. There were a few conditions where the motion that could not be categorized into the PPS motion, the PPC motion, the APC motion, or the SL motion. Such ambiguous motions, represented by cyan points in Fig.~\ref{fig4}, were observed at the edge of the region where the SL motion was observed (see Fig.~\ref{fig6} in Sect.~\ref{sec:discussion} for the details in these motions).

\begin{figure}[t]
\centering
\includegraphics{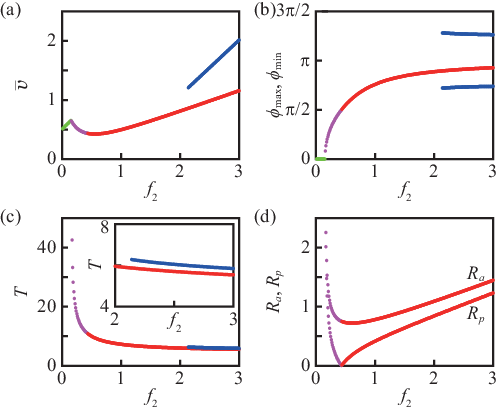}
\caption{(Color online) Dependence of (a) averaged COM speed $\bar{v}$, (b) absolute values of maximum and minimum angle differences $\phi_\mathrm{max}$ and $\phi_\mathrm{min}$ between $\bm{v}$ and $\bm{\ell}$, (c) period $T$ of the circular and slalom motions, and (d) radii of the trajectories of the active and passive particles, $R_a$ and $R_p$, for the circular motion depending on $f_2$. In panel (b), $\phi_\mathrm{max}$ and $\phi_\mathrm{min}$ coincide for the circular motion. The inset of panel (c) shows the expanded plot. In panel (d), the upper and lower plots correspond to $R_a$ and $R_p$, respectively. The color of the plot points correspond to the one in Fig.~\ref{fig4}. The parameters $f_1$ and $\eta$ were fixed at $f_1 = 0.5$ and $\eta = 0.5$, respectively. \label{fig5}}
\end{figure}

In order to clearly show the hysteresis of the stable states, the phase diagram with scanning $f_2$ was plotted in Fig.~\ref{fig5}, in which the averaged COM speed $\bar{v}$ [Fig.~\ref{fig5}(a)], the maximum and minimum values, $\phi_\mathrm{max}$ and  $\phi_\mathrm{min}$, of $\phi$ [Fig.~\ref{fig5}(b)], the period $T$ for the PPC, APC and SL motions [Fig.~\ref{fig5}(c)], and the radii of the circular trajectories of the active and passive particles for the PPC and APC motions [Fig.~\ref{fig5}(d)] are plotted. These values were obtained by scanning the value of $f_2$ in the upward and downward directions with the interval of $0.01$.

In Fig.~\ref{fig5}(b), the branch for the PPC motion seems to generate from the solution with $\phi = 0$ for the PPS motion. In addition, the averaged velocity is continuous between the PPS and PPC motions in Fig.~\ref{fig5}(a). Thus, the transition between the PPS and PPC motions is considered to be the supercritical pitchfork bifurcation, in which the symmetric solution with $\phi = 0$ becomes unstable. The boundary between the PPC and APC motions corresponds to the case with $R_p$ is zero as shown in Fig.~\ref{fig5}(d). This transition is not a type of bifurcation, but it can be understood that these two motions are connected by the motion in which the passive particle stops and active particle goes around it. The solution branches for APC and SL motions do not seem to be connected since we cannot find the point where the averaged COM speed or the period of motion coincides in Fig.~\ref{fig5}(a) and (c). The bifurcation structures related to these two motions should be clarified in future study.

\section{\label{sec:analysis}Linear stability analysis}

\subsection{Linearized equation}
We first consider the steady-state solutions of Eqs.~\eqref{eq_ell}, \eqref{eq_v}, and \eqref{eq_w}. Here we show the three types of steady-state solutions; the PPS motion, the active-particle preceding straight (APS) motion, and the circular motion. In the following subsection, we analyze the linear stability of these solutions. For the analyses, we introduce the components of each vector as $\bm{\ell} = \ell_x \bm{e}_x + \ell_y \bm{e}_y$, $\bm{v} = v_x \bm{e}_x + v_y \bm{e}_y$, and $\bm{w} = w_x \bm{e}_x + w_y \bm{e}_y$.

First, we obtain the linearized equations for Eqs.~\eqref{eq_ell}, \eqref{eq_v}, and \eqref{eq_w} around a fixed point $\bm{\ell}^* = \ell_x^* \bm{e}_x + \ell_y^* \bm{e}_y$, $\bm{v}^* = v_x^* \bm{e}_x + v_y^* \bm{e}_y$, and $\bm{w}^* = w_x^* \bm{e}_x +w_y^* \bm{e}_y$. By setting $\ell_x = \ell_x^* + \delta \ell_x$, $\ell_y = \ell_y^* + \delta \ell_y$, $v_x = v_x^* + \delta v_x$, $v_y = v_y^* + \delta v_y$, $w_x = w_x^* + \delta w_x$, and $w_y = w_y^* + \delta w_y$, the linearized equation is given as
\begin{align}
\frac{d}{dt} \begin{pmatrix}
    \delta \ell_x \\ \delta \ell_y \\ \delta v_x \\ \delta v_y \\ \delta w_x \\ \delta w_y
\end{pmatrix}
=
J \begin{pmatrix}
    \delta \ell_x \\ \delta \ell_y \\ \delta v_x \\ \delta v_y \\ \delta w_x \\ \delta w_y
\end{pmatrix},
\end{align}
where the Jacobian matrix $J$ is explicitly given as
\begin{widetext}
\begin{align}
 J = \begin{pmatrix}
        0 & 0 & 0 & 0 & 1 & 0 \\
        0 & 0 & 0 & 0 & 0 & 1 \\
        \frac{f_1 {\ell_y^*}^2}{2{\ell^*}^3} &
        -\frac{f_1 \ell_x^* \ell_y^*}{2{\ell^*}^3} & 
        \frac{f_2 {u_{y}^*}^2}{2 {u^*}^3} -\eta &
        -\frac{f_2 u_{x}^* u_{y}^*}{2 {u^*}^3} &
        -\frac{f_2 {u_{y}^*}^2}{4 {u^*}^3} &
        \frac{f_2 u_{x}^* u_{y}^*}{4 {u^*}^3} \\
        -\frac{f_1 \ell_x^* \ell_y^*}{2{\ell^*}^3} &
        \frac{f_1 {\ell_x^*}^2}{2{\ell^*}^3} &
        - \frac{f_2 u_{x}^* u_{y}^*}{2 {u^*}^3} &
        \frac{f_2 {u_{x}^*}^2}{2 {u^*}^3} -\eta &
        \frac{f_2 u_{x}^* u_{y}^*}{4 {u^*}^3} &
        -\frac{f_2 {u_{x}^*}^2}{4 {u^*}^3} \\       
        \frac{(f_1+2) {\ell_y^*}^2}{{\ell^*}^3}-2 &
        -\frac{(f_1+2) \ell_x^* \ell_y^*}{{\ell^*}^3} & 
        -\frac{f_2 {u_{y}^*}^2}{ {u^*}^3} &
        \frac{f_2 u_{x}^* u_{y}^*}{ {u^*}^3} &
        \frac{f_2 {u_{y}^*}^2}{2 {u^*}^3} - \eta &
        - \frac{f_2 u_{x}^* u_{y}^*}{2 {u^*}^3} \\
        -\frac{(f_1+2) \ell_x^* \ell_y^*}{{\ell^*}^3} &
        \frac{(f_1+2) {\ell_x^*}^2}{{\ell^*}^3}-2 &
        \frac{f_2 u_{x}^* u_{y}^*}{ {u^*}^3} &
        - \frac{f_2 {u_{x}^*}^2}{ {u^*}^3} &
        -\frac{f_2 u_{x}^* u_{y}^*}{2 {u^*}^3} &
        \frac{f_2 {u_{x}^*}^2}{2 {u^*}^3} - \eta
        \end{pmatrix}. \label{Jacobian}
\end{align}.
\end{widetext}
Here we set $\ell^* = \sqrt{{\ell_x^*}^2 + {\ell_y^*}^2}$, $\bm{u} = \bm{v}^* - \bm{w}^*/2 = u_x^* \bm{e}_x + u_y^* \bm{e}_y$, and $u^* = \sqrt{{u_{x}^*}^2 + {u_{y}^*}^2}$.  
Using these linearized equations, we discuss the linear stability for each solution below.

\subsection{Passive-particle preceding straight (PPS) motion}

First, we consider the solution corresponding to the PPS motion. Considering the symmetric properties, we can assume the motion is in the positive $x$-direction without losing generality. The fixed point corresponding to the steady-state solution is given as
\begin{align}
\bm{\ell}^* = \ell_0^{(\mathrm{p})} \bm{e}_x,
\end{align}
\begin{align}
\bm{v}^* = v_0^{(\mathrm{p})} \bm{e}_x,
\end{align}
\begin{align}
\bm{w}^* = \bm{0} ,
\end{align}
where 
\begin{align}
\ell_0^{(\mathrm{p})} = \frac{f_1 - f_2}{2} + 1,
\end{align}
and
\begin{align}
v_0^{(\mathrm{p})} = \frac{f_1 + f_2}{2 \eta}.
\end{align}

Now we consider the PPS motion in the positive $x$-direction, and thus $v_0^{(\mathrm{p})}$ and $\ell_0^{(\mathrm{p})}$ should be positive. Therefore, $f_1 + 2 > f_2$ should hold.

By substituting these solutions into the Jacobian matrix $J^{(\mathrm{p})}$ of the linearized equation in Eq.~\eqref{Jacobian}, we obtain
\begin{align}
    J^{(\mathrm{p})} = \begin{pmatrix}
        0 & 0 & 0 & 0 & 1 & 0 \\
        0 & 0 & 0 & 0 & 0 & 1 \\
        0 & 0 & -\eta & 0 & 0 & 0 \\
        0 & \frac{f_1}{f_1 - f_2 +2} & 0 & -\frac{\eta f_1}{f_1 + f_2} & 0 & -\frac{\eta f_2}{2(f_1 +f_2)} \\
        -2 & 0 & 0 & 0 & -\eta & 0 \\
        0 & \frac{2f_2}{f_1 - f_2 + 2} & 0 & - \frac{2 \eta f_2}{f_1 + f_2} & 0&  - \frac{\eta f_1}{f_1 + f_2}
        \end{pmatrix}.
\end{align}
The eigenvalues $\lambda^{(\mathrm{p})}$ of $J^{(\mathrm{p})}$ are given as
\begin{align} 
\lambda^{(\mathrm{p})} =& 0, -\eta, \frac{-\eta \pm \sqrt{\eta^2 - 8}}{2}, \nonumber\\
     &-\eta + \frac{f_2}{2v^{(\mathrm{p})}_0} \pm \frac{1}{2} \sqrt{\frac{{f_2}^2}{{v^{(\mathrm{p})}_0}^2} + \frac{4f_2}{\ell^{(\mathrm{p})}_0} }.
\end{align}
Since the real parts of $-\eta$ and $\left(-\eta \pm \sqrt{\eta^2 - 8} \right)/2$ are always negative, the signs of the real parts of the last eigenvalues determine the linear stability of the motion. Considering that the part inside the square root is always positive, the eigenvalue with the negative sign is always negative. The one with the positive sign can change its sign; it is negative when
\begin{align}
    \eta^2  \frac{f_1 - f_2}{f_1 + f_2} >  \frac{f_2}{\ell_0^{(\mathrm{p})}} = \frac{2 f_2}{f_1 - f_2 + 2}
\end{align}
holds by considering that $-\eta + f_2/(2v_0^{(\mathrm{p})}) = -\eta f_1/(f_1 + f_2) <0$. Since the above equation holds only when $f_1 > f_2$, we finally obtain the condition for linear stability as
\begin{align}
\eta > \sqrt{ \frac{2 f_2 (f_1  +f_2)}{(f_1 - f_2)(f_1 - f_2  +2 )} }. \label{line_th}
\end{align}
This suggests that the pitchfork bifurcation occurs at $\eta = \sqrt{2f_2(f_1 + f_2)/((f_1 - f_2)(f_1 - f_2 + 2))}$. Taken the numerical simulation results into consideration, it seems to be supercritical.
It should be noted that the eigenvector with respect to the zero eigenvalue is ${}^\mathrm{t}(0, \ell_0^{(\mathrm{p})}, 0 ,v_0^{(\mathrm{p})}, 0, 0)$, which corresponds to the isotropy of the system.

\subsection{Active-particle preceding straight (APS) motion}

Next, we consider the solution corresponding to the APS motion, which is not realized as a stable solution in the numerical simulations. In the same manner as that for the solution corresponding to the PPS motion, we can assume the motion is in the positive $x$-direction without losing generality. The fixed point corresponding to the steady-state solution is given as
\begin{align}
\bm{\ell}^* = -\ell_0^{(\mathrm{a})} \bm{e}_x,
\end{align}
\begin{align}
\bm{v}^* = v_0^{(\mathrm{a})} \bm{e}_x,
\end{align}
\begin{align}
\bm{w}^* = \bm{0},
\end{align}
where 
\begin{align}
v_0^{(\mathrm{a})} = \frac{-f_1 + f_2}{2 \eta} ,
\end{align}
and
\begin{align}
\ell_0^{(\mathrm{a})} = \frac{f_1 + f_2}{2} + 1.
\end{align}
Here, we consider the APS motion, and thus $v_0^{(\mathrm{a})}$ and $\ell_0^{(\mathrm{a})}$ should be positive. Therefore, $f_2 > f_1$ should hold.

By substituting these solutions into the Jacobian matrix $J^{(\mathrm{a})}$ of the linearized equation in Eq.~\eqref{Jacobian}, we obtain
\begin{align}
    J^{(\mathrm{a})} = \begin{pmatrix}
        0 & 0 & 0 & 0 & 1 & 0 \\
        0 & 0 & 0 & 0 & 0 & 1 \\
        0 & 0 & -\eta & 0 & 0 & 0 \\
        0 & \frac{f_1}{f_1 + f_2 +2} & 0 & \frac{\eta f_1}{f_2 - f_1} & 0 & -\frac{\eta f_2}{2(f_2 - f_1)} \\
        -2 & 0 & 0 & 0 & -\eta & 0 \\
        0 & -\frac{2f_2}{f_1 + f_2 + 2} & 0 & -\frac{2 \eta f_2}{f_2 - f_1} & 0&  \frac{\eta f_1}{f_2 - f_1}
        \end{pmatrix}.
\end{align}
The eigenvalues $\lambda^{(\mathrm{a})}$ of $J^{(\mathrm{a})}$ are given as
\begin{align}
\lambda^{(\mathrm{a})} =& 0, -\eta, \frac{-\eta \pm \sqrt{\eta^2 - 8}}{2}, \nonumber\\
     &-\eta + \frac{f_2}{2v^{(\mathrm{a})}_0} \pm \frac{1}{2} \sqrt{\frac{{f_2}^2}{{v^{(\mathrm{a})}_0}^2} - \frac{4f_2}{\ell_0^{(\mathrm{a})}} }.
\end{align}
It should be also noted that the eigenvector with respect to the zero eigenvalue is ${}^\mathrm{t}(0, -\ell_0^{(\mathrm{a})}, 0 ,v_0^{(\mathrm{a})}, 0, 0)$, which corresponds to the isotropy of the system.
In the same manner as that in the case of the PPS motion, the real parts of $-\eta$ and $\left(-\eta \pm \sqrt{\eta^2 - 8} \right)/2$ are always negative. In contrast, one of the real parts of the last eigenvalues are 
always positive since $-\eta + f_2/(2v_0^{(\mathrm{a})}) = f_1 /(\eta(f_2- f_1)) > 0$. 
The term inside the square root is negative when
$\eta < \left|f_2-f_1\right| \sqrt{2 / (f_2 (f_1 + f_2 + 2))}$. 
In this case, the last eigenvalues are complex conjugates whose imaginary parts are $i\sqrt{f_2\left[2/(f_1 + f_2 + 2) - f_2 \eta^2 / (f_2 - f_1)^2\right]} \equiv i \Omega$.

In the limit of $f_2 \to +\infty$ with fixed $f_1$ and $\eta$, the real part $-\eta + \eta f_2/(f_2 - f_1)$ approaches 0. This suggests the oscillatory behavior with a period of $2\pi/\Omega$ around the fixed point for large $f_2$. We presume that this oscillation corresponds to the SL motion observed in the numerical simulation. That is to say, the SL motion appears through the oscillatory destabilization of the traveling direction in the APS motion. Actually, the angle $\phi$ between $\bm{\ell}$ and $\bm{v}$ oscillates around $\pi$ as shown in Fig.~\ref{fig5}(b). Considering that $\Omega$ approaches $\sqrt{2 - \eta^2}$ in the limit of $f_2 \to \infty$, the corresponding period for $\eta = 0.5$ is calculated as around $4.75$. This value is slightly smaller than the period obtained by the simulation results in Fig.~\ref{fig5}(c). The theoretical approach gives the value with the same order, and the slight difference seems to be due to the finite amplitude of the SL motion.

\subsection{Circular motion}

In order to discuss the existence and stability of a circular motion, we consider the transform to the polar coordinates $(r, \theta)$ in which the origin meets the center of the circular orbit.
The unit vectors in the polar coordinates are set as $\bm{e}_r = \cos \theta \bm{e}_x + \sin \theta \bm{e}_y$ and $\bm{e}_\theta = -\sin\theta \bm{e}_x + \cos \theta \bm{e}_y$. Then, we set the variables in the polar coordinates as
\begin{align}
    \bm{r} = r\bm{e}_r,
\end{align}
\begin{align}
    \bm{v} = v \left( \cos \xi \bm{e}_r + \sin \xi \bm{e}_\theta \right),
\end{align}
\begin{align}
    \bm{\ell} = \ell \left[\cos (\phi + \xi) \bm{e}_r + \sin (\phi + \xi) \right]\bm{e}_\phi,
\end{align}
\begin{align}
    \bm{v} - \frac{\bm{w}}{2} = u \left[ \cos (\psi + \xi) \bm{e}_r + \sin (\psi + \xi) \bm{e}_\theta \right],
\end{align}
Here, we adopt $\bm{u} = \bm{v} - \bm{w}/2 = \bm{v}_a$ since the equations become simple by this transform. We also adopt the relative angles $\phi + \xi$ and $\psi + \xi$ for $\bm{\ell}$ and $\bm{u}$, respectively, for simple expressions. By careful calculation from Eqs.~\eqref{eq_r}--\eqref{eq_w}, we obtain the dynamical systems with respect to 8 variables, $r$, $\theta$, $v$, $\xi$, $\ell$, $\phi$, $u$, and $\psi$ as
\begin{align}
\frac{dr}{dt} = v \cos \xi, \label{eq_r_pol}
\end{align}
\begin{align}
\frac{d\theta}{dt} =& \frac{v}{r} \sin \xi,  \label{eq_theta_pol}
\end{align}
\begin{align}
\frac{d\ell}{dt} = 2 v \cos \phi - 2 u \cos (\phi - \xi), \label{eq_ell_pol}
\end{align}
\begin{align}
\frac{d\phi}{dt} = - \left ( \frac{2v}{\ell} + \frac{f_1}{2v} \right ) \sin \phi - \frac{f_2}{2v} \sin \psi + \frac{2u}{\ell} \sin (\phi - \psi), \label{eq_phi_pol}
\end{align}
\begin{align}
\frac{dv}{dt} = - \eta v + \frac{f_1}{2} \cos \phi + \frac{f_2}{2} \cos \psi , \label{eq_v_pol}
\end{align}
\begin{align}
\frac{d\xi}{dt} = - \frac{v}{r} \sin \xi + \frac{f_1}{2v} \sin \phi + \frac{f_2}{2v} \sin \psi, \label{eq_zeta_pol}
\end{align}
\begin{align}
\frac{du}{dt} = - \eta u + ( \ell - 1) \cos (\phi - \psi) + f_2, \label{eq_u_pol}
\end{align}
\begin{align}
\frac{d\psi}{dt} = - \frac{f_1}{2v} \sin \phi - \frac{f_2}{2v} \sin \psi + \frac{\ell - 1}{u} \sin (\phi - \psi). \label{eq_psi_pol}
\end{align}
Since the three variables $r$, $\theta$, and $\xi$ do not appear in the left sides of the equations \eqref{eq_ell_pol}, \eqref{eq_phi_pol}, \eqref{eq_v_pol}, \eqref{eq_u_pol}, and \eqref{eq_psi_pol}, the dynamical system is essentially regarded as a 5-variable one.

If we consider a circular motion in which the center of orbit meets the origin, only the variable $\theta$ depends on the time $t$ as $\theta = \omega t$, where $\omega$ is the angular velocity, and the other variables are constant. Therefore, the circular motion corresponds to the fixed points of the 5-variable dynamical systems.

The linearized equations of Eqs.~\eqref{eq_ell_pol}, \eqref{eq_phi_pol}, \eqref{eq_v_pol}, \eqref{eq_u_pol}, and \eqref{eq_psi_pol} around a fixed point $\ell^*$, $\phi^*$, $v^*$, $u^*$, and $\psi^*$ are given, by setting $\ell = \ell^* + \delta \ell$, $\phi = \phi^* + \delta \phi$, $v = v^* + \delta v$, $u = u^* + \delta u$, and $\psi = \psi^* + \delta \psi$, as
\begin{align}
\frac{d}{dt} \begin{pmatrix}
    \delta \ell \\ \delta \phi \\ \delta v \\ \delta u \\ \delta \psi
\end{pmatrix}
=
\tilde{J} \begin{pmatrix}
   \delta \ell \\ \delta\phi \\ \delta v \\ \delta u \\ \delta \psi
\end{pmatrix},
\end{align}
where Jacobian matrix $\tilde{J}$ is explicitly given as
\begin{widetext}
\begin{align}
    \tilde{J} = \begin{pmatrix}
        0 & \frac{\ell^*(\ell^*-1) \sin \zeta^*}{\ell^*} & 2 \cos \phi^* & -2 \cos \zeta^* & -2u \sin \zeta^* \\
        \frac{-(\ell-1)\sin \zeta^*}{\ell^* u^*} & -\frac{f_1\cos\phi^*}{2v^*} & \frac{(\ell^*-1) \sin \zeta^*}{u^* v^*} -\frac{2 \sin\phi^*}{\ell^*} & \frac{2\sin \zeta^*}{\ell^*} & \frac{- 2u^* \cos\zeta^*}{\ell^*} - \frac{f_2 \cos \psi^*}{2} \\
        0 & -\frac{ f_1 \ell^*\sin \phi^*}{2} & -\eta & 0 & - \frac{l^* f_2 \sin\psi^*}{2} \\
        \cos\zeta^* & -(\ell^* -1)\sin \zeta^*  & 0 & -\eta & (\ell^* -1)\sin \zeta^* \\
        \frac{\sin \zeta^*}{u^*} & \frac{-(l^*-1)\cos\zeta^*}{u^*} - \frac{f_1 \cos\phi^*}{2v^*} & \frac{(\ell^*-1 )\sin \zeta^*}{u^* v^*} & \frac{-(l^*-1) \sin\zeta^*}{{u^*}^2} & \frac{-(l^*-1) \cos\zeta^*}{u^*} - \frac{f_2 \cos \psi^*}{2v^*}
        \end{pmatrix}, \label{Jacobian_p}
\end{align}
\end{widetext}
where we define $\zeta^* = \phi^* - \psi^*$.

We numerically obtained the fixed point for each parameter and calculated the eigenvalues of $\tilde{J}$. Then, the real parts of all eigenvalues were negative, and therefore it was concluded that the solution corresponding to the circular motion is asymptotically stable.

\section{\label{sec:discussion}Discussion}

From the simulation results, the pair of active and passive particles interacting through chemo-repellant concentration field exhibits a passive-particle preceding straight (PPS) motion, a passive-particle preceding circular (PPC) motion, an active-particle preceding circular (APC) motion, or a slalom (SL) motion. From the theoretical analysis, the transition between the PPS and PPC motions is identified to be a pitchfork bifurcation
as we consider $f_2$ as a bifurcation parameter. As for the SL motion, the solution branch corresponding to the SL motion does not connect to the other three solution branches. We consider that the SL motion appears through the instability of the active-particle preceding straight (APS) motion, which is always unstable. 

\begin{figure}
\centering
\includegraphics{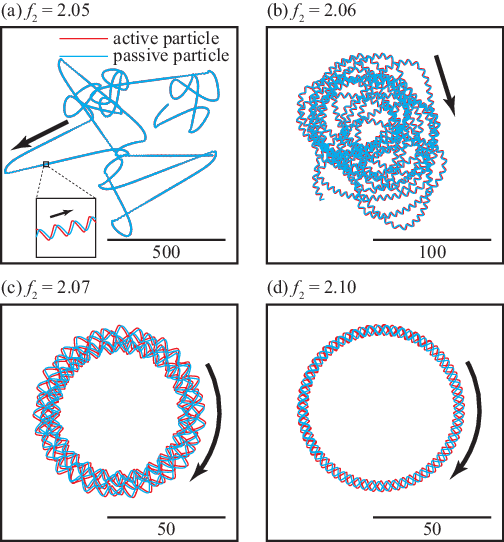}
\caption{(Color online) Trajectories of the active and passive particles after sufficient time. (a) $f_2 = 2.05$. (b) $f_2 = 2.06$. (c) $f_2 = 2.07$. (d) $f_2 = 2.10$. Red and cyan curves show the trajectories of the active and passive particles, respectively. The parameters $f_1$ and $\eta$ were fixed at $f_1 = 0.5$ and $\eta = 0.5$, respectively. \label{fig6}}
\end{figure}

In Fig.~\ref{fig4}, there are several points corresponding to the ambiguous motion at the edge of the region for the bistability of the SL and APC motions. We have checked the dynamics by precisely changing the value of $f_2$, where the calculation method was the same as Fig.~\ref{fig2}. The trajectories of active and passive particles are shown in Fig.~\ref{fig6},  where $f_2$ was set to be 2.05 (a), 2.06 (b), 2.07 (c), and 2.10 (d). As the parameter $f_2$ decreased, the SL motion changed to the slalom motion along a circle as shown in Fig.~\ref{fig6}(d), and then to the circular slalom motion along a curve like the Lissajous figure as shown in Fig.~\ref{fig6}(c). With a further decrease in $f_2$, the chaotic motion was observed as shown in Fig.~\ref{fig6}(a,b). In this case, the curve along which the slalom motion was exhibited seemed to be chaotic. The detailed analyses of these series of bifurcation structure may be interesting from the viewpoint of dynamical systems, though we leave them as future study.

In our previous study~\cite{Ishikawa_PRE}, we investigated a pairing-induced motion of a camphor particle and a metal washer at a water surface. Our model was constructed motivated by the experiments, where a camphor disk (active particle) moves by the self-phoresis, in which the particle moves in the direction of concentration gradient of released camphor molecules. Due to this effect, the metal washer (passive Particle) moves away from the active particle and the active particle is driven in the direction of the velocity of itself. Moreover, the active and passive particles interact with each other through the lateral capillary force~\cite{CHAN1981410}. In our present model, we consider a harmonic potential by a spring between the active and passive particles since the lateral capillary force only plays a role in connecting the two particles. Since the distance between the two particles is almost kept constant, we assume that the magnitude of the repulsive force working on the passive particle is constant. As for the magnitude of the force working on the active particle, we also set it constant since the particle is moving at a constant velocity. Considering that the camphor particle releases camphor molecules and reduces the surface tension, Marangoni flow should be induced. Moreover, the drag force originating from the hydrodynamic flow structure around the accompanying two particles should work. The effect by Marangoni flow is included in the force working on the active particle, while the drag force is introduced in the harmonic potential by a spring between the active and passive particles. In our model, the characteristic length between the particles is on the same order of the diffusion length~\cite{suematsu2014quantitative,JCP_HK_NY}. Considering that both lengths are of the order of 10~mm in the experiments, the parameter values adopted in the numerical simulation are reasonable. It should be noted that this assumption does not hold when a particle moves slowly since the asymmetry in concentration field around the particle becomes small. Especially, our model fails when the active particle stops. However, we do not need to mind this since the active particle moves as long as the passive particle is located close to the active particle.

In the experiment reported in our previous study~\cite{Ishikawa_PRE}, the pair of the camphor disk (active particle) and metal washer (passive particle) show the straight motion in which the passive particle precedes when the viscosity of the aqueous solution was high, while it showed the circular motion when the viscosity was low. The difference in the viscosity can be represented by varying $\eta$ in our present model. Numerical simulation results using our present model show that the particle pair bifurcates from the PPS motion to a circular motion with a decrease in $\eta$. This well corresponds to the experimental result. So far, the slalom motion was not observed in the experiment. This seems to be because the spring connecting between the two particles does not properly correspond to the experimental system; the harmonic potential is introduced to simply describe the effect of the lateral capillary force, though the lateral capillary force only works as an attractive force, but not as a repulsive force. Our results suggest that the pair of active and passive particles may exhibit a slalom motion if they have the interaction with a preferable distance. 

Finally, we may also emphasize that our model holds for a general system composed of interacting self-phoretic particles releasing chemo-repellant ~\cite{mikhailov2013cells,Grima_PRL,Dean_PRE2024}. We hope our study will motivate researchers to build actual experimental systems.

\section{\label{sec:summary}Summary}

In the present study, we constructed a mathematical model for the pairing-induced motion of active and passive particles, which was motivated from the experiments using a camphor disk and a metal washer floating on water~\cite{Ishikawa_PRE}.
The model was simplified by assuming that the forces working on the active and passive particles due to the chemo-repellant concentration field have the constant absolute value and only depend on the configuration and velocities of the particles. By numerical simulation, we found the four types of motions, the passive-particle preceding straight (PPS), passive-particle preceding circular (PPC), active-particle preceding circular (APC), and slalom (SL) motions. We also performed a linear stability analysis and clarified that the transition between the PPS and PPC motions is considered to be a supercritical pitchfork bifurcation, which well corresponds to the experimental results. We also discussed the mechanism of the SL motion related to the instability of the active-particle preceding straight (APS) motion. We believe our present model will be a good candidate for a fundamental model for investigating multiple particle systems with the mixture of active and passive particles by virtue of its simple setup.  

\begin{acknowledgments}
This work was supported by JSPS KAKENHI Grants Nos. JP20H02712, JP21H00996, JP21H01004, and JP24K16981 and by the Cooperative Research Program of ``Network Joint Research Center for Materials and Devices'' (Nos.~20241063 and 20244003). This work was also supported by PAN-JSPS program (No.~JPJSBP120234601) and by the JSPS Core-to-Core Program ``Advanced core-to-core network for the physics of self-organizing active matter'' (JPJSCCA20230002).
\end{acknowledgments}

\end{document}